\documentclass[%
 reprint,
 amsmath,amssymb,
 aps,
 pra,
]{revtex4-2}

\usepackage{graphicx}% Include figure files
\usepackage{dcolumn}% Align table columns on decimal point
\usepackage{bm}% bold math

\begin{document}

\preprint{APS/123-QED}

\title{Limitations in Fluorescence-Detected Entangled Two-Photon-Absorption Experiments: Exploring the Low- to High-Gain Squeezing Regimes}% Force line breaks with \\
\author{Tiemo Landes}
\author{Brian J. Smith}%
\author{Michael G. Raymer}%
\affiliation{Department of Physics and Oregon Center for Optical, Molecular, and Quantum Science, University of Oregon, Eugene, Oregon 97403, USA}%Lines break automatically or can be forced with \\
 \email{raymer@uoregon.edu}
\date{\today}

\begin{abstract}
We closely replicated and extended a recent experiment (“Spatial properties of entangled two-photon absorption,” Phys. Rev. Lett. 129, 183601, 2022) that reportedly observed enhancement of two-photon absorption rates in molecular samples by using time-frequency-entangled photon pairs, and we found that in the low-flux regime, where such enhancement is theoretically predicted in-principle, the two-photon fluorescence signal is below detection threshold using current state-of-the-art methods. The results are important in the context of efforts to enable quantum-enhanced molecular spectroscopy and imaging at ultra-low optical flux. Using an optical parametric down-conversion photon-pair source that can be varied from  the low-gain spontaneous regime to the high-gain squeezing regime, we observed two-photon-induced fluorescence in the high-gain regime but in the low-gain regime any fluorescence was below detection threshold. We supplemented the molecular fluorescence experiments with a study of nonlinear-optical sum-frequency generation, for which we are able to observe the low-to-high-gain crossover, thereby verifying our theoretical models and experimental techniques. The observed rates (or lack thereof) in both experiments are consistent with theoretical predictions and with our previous experiments, and indicate that time-frequency photon entanglement does not provide a practical means to enhance in-solution molecular two-photon fluorescence spectroscopy or imaging with current techniques.
\end{abstract}

\maketitle

\section{Introduction}

The use of the spectral-temporal correlation present in time-frequency entangled photon pairs to drive nonlinear optical light-matter interactions more efficiently than can be achieved with classical laser light has been studied since first being proposed in 1989 \cite{Gea-Banacloche1989, Javanainen1990}. Promising initial work utilizing squeezed light driving narrow-line atomic systems was demonstrated in the mid 1990s and 2000s \cite{Georgiades1995, dayan2004}. Recently, the possibility of so-called entangled two-photon absorption (ETPA) in molecular samples of chemical and biological interest has been the subject of lively debate. Studies dating back to 2006 \cite{Goodson2006,Villabona-Monsalve2017,Goodson2017,Tabakaev2020} claim to observe ETPA in molecular systems at efficiencies estimated to be around 10 orders of magnitude larger than expected from heuristic theoretical arguments \cite{saleh1998}. More recent work has demonstrated functional microscopes purportedly benefiting from time-frequency entanglement of photons  \cite{Goodson2020Varnavski, Varnavski2022}. However, the magnitude of the reported increase in two-photon absorption (TPA) efficiency by use  of photonic entanglement cannot be easily explained by standard quantum mechanical models, which indicate that any enhancement should be far more moderate than reported  \cite{Dayan2007,raymer2021memo,Landes2021Opex,drago2022}. The estimates presented by these theories indicate that measurements of ETPA would be infeasible with currently available entangled-photon sources based on spontaneous parametric down conversion, two-photon absorbing samples and current detection methods. Recent studies have placed experimental upper bounds on the efficiency of ETPA\cite{parzuchowski2020,Landes2021PRR} under various experimental conditions that are consistent with the theoretical estimates. Separate experiments have described other linear-optical effects that could be mistaken for ETPA using common experimental methods to search for ETPA signals\cite{Mikhaylov2022, Cushing2022, Acquino2022}.  

In this work, we present a series of experiments examining TPA in Rhodamine 6G (R6G) molecular dye in solution and, for comparison, the closely-related nonlinear optical process of sum-frequency generation (SFG) using various excitation sources. The TPA experiments aimed to reproduce results from a recent study that reports ETPA in R6G solution \cite{tabakaev2022} while placing tight constraints on the source and detection systems. Our experiment had several key improvements over the original, as described in the methods sections. In brief, they are: a lower-noise detection system, a flexible laser system allowing observation of TPA-induced fluorescence using light  generated by high-gain squeezing, and the observation of the low-to-high-gain 'crossover' in SFG to check consistency of our experimental setup with theory predictions. The main result is that using this well-calibrated system, no detectable ETPA was observed, casting doubt or at least questions about the origin of the signals reported elsewhere. 

Here quantum light is generated by multimode spontaneous parametric down conversion (SPDC) using both continuous-wave and pulsed-pump configurations. The SPDC process generates squeezed vacuum states of light going from the low- to high-gain regimes to observe the scaling of TPA and SFG signals arising from these different input states of light. In the limit of low-gain squeezing, where the quantum-advantage of using EPP is predicted to be most pronounced \cite{Dayan2007,raymer2021memo,Landes2021Opex,drago2022}, we can access the so-called 'isolated-pair regime' in which there is at most one pair of photons distributed across the correlated source modes. The enhancement in TPA efficiency arising from spectral-temporal correlations of the source is expected to occur when the spectral bandwidth of the entangled photon pair is broader than the linewidth of the two-photon absorption transition. The pulsed version of the experiment enables us to clearly observe TPA using a high flux of squeezed light. We verified the presence of the TPA process in this case by studying the fluorescence emission lineshape and quadratic dependence of fluorescence intensity with excitation flux. We use squeezed-vacuum light in the high-gain regime, also called bright squeezed vacuum (BSV) to probe the efficiency of TPA and determine the minimum flux required to observe fluorescence from TPA in the pulsed configuration. Separately, we use the pulsed source to confirm the ability of the standard quantum optical model to accurately predict the properties of sum-frequency generation (SFG) of squeezed vacuum, which shares many similarities to TPA and confirm important predictions about the process driven by squeezed light \cite{Dayan2007}.

With careful calibration of the source and detection system, we were unable to observe any TPA fluorescence signal when using weak squeezed vacuum (isolated photon pairs) [20], consistent with predictions of recent calculations of ETPA efficiency\cite{raymer2021memo,Landes2021Opex}. These calculations predict only a moderate ETPA efficiency as a result of the exceedingly small TPA cross sections in solvated dye molecules. The predicted efficiency is insufficient to outweigh the low flux required to operate in the isolated entangled-photon-pair regime where enhancement by entanglement is predicted, rather than in the BSV regime where enhancement by entanglement is predicted to be minimal or nonexistent. The present experiments provide evidence that the signals attributed to ETPA reported in previous studies in molecules, in which the TPA linewidth is large compared to the EPP bandwidth, likely originated from unknown mechanisms, not clearly related to ETPA. Thus, the present study provides strong evidence that the sought-after advantages of using ETPA as a practical tool for enhancing molecular spectroscopy and biological imaging remains elusive. 

This paper first brings together and summarizes relevant theoretical considerations, then presents the results of our experimental study.

\section{Theory summary}

The scenario considered is two-photon excitation of a molecular system in solution by a broadband squeezed-vacuum field having a single linear polarization and with entanglement in frequency (anti-correlated) and in time (positively correlated), as commonly generated using spontaneous parametric down-conversion (SPDC).

The theory of two-photon absorption of time-frequency entangled photons for atomic or molecular systems with no near-resonant intermediate states has been described in detail \cite{Dayan2007, raymer2021tutorial}. While the dyes used in our experiments and our theoretical models correspond to TPA without near-resonant intermediate states, theoretical work on systems with intermediate states has also been reported, reaching similar conclusions\cite{drago2022, Dorfman2016}. Our experiments range from the low-gain SPDC regime to the regimes of intermediate and high-gain squeezed vacuum. To distinguish between the regimes we refer to low-gain squeezed vacuum as the isolated-pair regime in which entangled photon pairs (EPP) are generated and the probability for multi-pair emission is small. We refer to the high-gain squeezed vacuum regime as bright squeezed vacuum (BSV)\cite{Spasibko2017_checkhova_g2g3g4}. When referring to both the low- and high-gain regimes we simply use the term squeezed vacuum. At intermediate gains, in which the multi-pair emission becomes comparable to the single-pair emission, a ‘crossover’ regime in the scaling of the TPA rate is explored. These are different regimes of the same squeezing process, and are generated in the same nonlinear optical crystal in our experiments.

Theoretical derivations of the TPA rate point to a positive enhancement (compared with laser light with the same spectral bandwidth and flux) of the efficiency of TPA of squeezed vacuum in the isolated-pair regime as a result of time-frequency correlations and photon-number correlations   \cite{raymer2021tutorial, Dayan2007,raymer2021memo,Landes2021Opex,raymer2022}. This relative increase in TPA efficiency can be large, but only in the case of ultra-low entangled-pair flux and for narrow molecular or atomic TPA linewidths compared with the bandwidth of the excitation source. However, due to the extremely low efficiency of the TPA process itself and challenges in realizing a high generation rate of isolated entangled photon pairs, the net ETPA rates are predicted theoretically to be well below current detection thresholds for common molecular dyes and typical experimental geometries \cite{Landes2021Opex}.

\subsection{Linear and Quadratic Flux Scaling}

A key experimental feature of ETPA is the predicted linear scaling of the EPTA rate with excitation flux, not expected for most second-order nonlinear optical phenomena, which typically display quadratic scaling for classically driven processes. This linear flux scaling feature alone is cited as evidence of ETPA in many experiments that have observed signals attributed to ETPA\cite{Goodson2006, Villabona-Monsalve2017,Tabakaev2020}.

This linear scaling can be understood in terms of a commonly adopted heuristic equation for the rate of ETPA for a single absorber (atom or molecule)\cite{saleh1998} 
\begin{equation}
    R^{ETPA} = \sigma_e \phi+\xi\sigma^{(2)}\phi^2,
    \label{ETPA Rate}
\end{equation}
\noindent where $\phi$ is the photon-flux density [$\text{photons m}^{-2} \text{s}^{-1}$], $\sigma_e$ is the ETPA cross-section  [$\text{m}^{2}$],$\sigma^{(2)}$ is the conventional TPA cross section [GM]. We have also inserted $\xi$ which is a unitless parameter of order unity discussed more below. The first term arises from a coherent contribution in which all photon pairs contributing to TPA are temporally-spectrally correlated, the second term arises from both correlated photons pairs and uncorrelated photon pairs. We refer to the contributions from the uncorrelated pairs as the incoherent contribution. 

For molecules in solution $\sigma^{(2)}$ is typically exceedingly small—on the order of 1 to 1,000 GM, where 1 GM = $10^{-58} \text{m}^{4} \text{s}$ \cite{BOYD2008}. When the bandwidth of the exciting squeezed-state light is much greater than the molecular TPA linewidth, $\sigma_e$ is given by
\begin{equation}
    \sigma_e=f \sigma^{(2)}/A_eT_e,
    \label{sigmae}
\end{equation}
\noindent where $A_e$ and $T_e$ are the entanglement area and entanglement time  of the exciting light (defined as the area and time within which photon pairs are tightly correlated), and $f$ is a phenomenological parameter. This heuristic description agrees well with in-depth theoretical derivations in cases where the squeezed light is confined to a single transverse mode (such that $A_e$ equals the transverse mode area), which show clearly that $f$ is expected to be of the order of unity \cite{raymer2022,Dayan2007}. 

In contrast to ETPA, TPA driven by coherent or thermal-like light lacks the linear-in-flux term because in those cases the probability for two photons to arrive within the molecular response time depends quadratically on the incident flux. For ETPA the $\sigma_e$ term arises due to the fact that  entangled photons are produced in pairs, leading in the low-flux limit to a linear relationship with the photon flux. 

The in-depth derivations also show that when the bandwidth of the exciting squeezed-state light is much smaller than the molecular TPA linewidth, meaning the two-photon response is instantaneous, the rate is given by \cite{raymer2022},
\begin{equation}
    R^{ETPA} = g^{(2)}(0)\sigma^{(2)} \phi^2,
    \label{ETPA g2}
\end{equation}
\noindent where $g^{(2)}(0)$ is the degree of second-order coherence of the light. In the case of type-0 or type-I SPDC phase matching--where the photons are indistinguishable--it is given by $g^{(2)}(0)=1/n +3$, in which $n$ is the mean number of photons per spectral-temporal-spatial mode \cite{Dayan2007,Dorfman2016}. In this case the factor $\xi$ in the heuristic equation is equal to 3 in the high-flux regime. For type-II phase matching, wherein photons are distinguishable, $g^{(2)}(0)=1/n +2$ \cite{raymer2022}. Note that this equation still maintains linear flux dependence at low gain due to the implicit dependence of $g^{(2)}(0)$ on $1/\phi$ , since $\phi \propto n$.

 In cases where the squeezed light is confined to a single transverse mode, such that $A_e$ equals the transverse mode area $A_0$, we have $n=\phi A_0 T_e$, where the entanglement time $T_e$ equals the inverse of the full-width-at-half-maximum (FWHM) bandwidth B (in Hz) of the squeezed-light source $T_e=1/B$, assuming the SPDC pump laser bandwidth is much smaller than the SPDC bandwidth, $\Delta \nu _{p} << B$. The case when the bandwidth of the squeezed light source becomes comparable to the molecular TPA linewidth is described by more complicated formulas, as explored in \cite{raymer2021tutorial, raymer2022}. 

If linear loss, e.g., an optical attenuator, is inserted between the SPDC source and the TPA sample, the flux scaling of the TPA rate in the low-flux regime is predicted to be quadratic with the amount of loss, because both photons in a pair are independently subject to the loss. This feature has been verified in sum-frequency generation (SFG) \cite{Dayan2005_SFG, Landes2021PRR}, which is a separate nonlinear optical process with  dynamics similar to that of TPA in the absence of intermediate resonances \cite{Dayan2007, raymer2022}. This feature is also readily observable in photon coincidence counting experiments, which are insensitive to ultra-fast temporal or spectral effects. Observation of both the linear behavior (when varying the SPDC pump power) and quadratic behavior (when inserting loss) in the same experiment is a more robust validation than verifying linear scaling alone, since linear scaling in itself is a common phenomenon and can lead to mis-attribution of unrelated linear-optical effects to ETPA \cite{Mikhaylov2022, Cushing2022, Acquino2022}. To date, only \cite{tabakaev2022} has reported both effects simultaneously in molecular TPA.

It is worth noting that, as pointed out in \cite{tabakaev2022}, $\sigma_e$ is a quantity that depends on the molecular properties of the TPA sample as well as the detailed properties of the entangled light driving the interaction. This includes experimental conditions such as beam focusing, pump bandwidth, as well as the phase-matching bandwidth of the process generating the entangled photons. Because of this, it is ambiguous to report $\sigma_e$ without explicit consideration of these parameters. A modified quantity $\sigma_e\times T_e\times A_e$ is more physically meaningful for comparison of experiments with varying conditions. 

The primary question we wish to address is what values can the parameter $f$ take on in realistic experiments? While theory indicates that $f$ should be on the order of 1, \cite{Landes2021Opex} the value needed to produce the ETPA efficiencies reported in other ETPA experiments\cite{Goodson2006,tabakaev2022} is many orders of magnitude greater. The present paper reports further concrete experimental evidence that $f$ is not much greater than $1$ in solvated dye-like molecules.

\subsection{Crossover Flux}

The crossover between the linear and quadratic flux-scaling regimes of two-photon absorption  and SFG rates for a narrowband two-photon absorbing sample occurs when the two terms in Eq. (\ref{ETPA Rate}) are comparable, as described in \cite{Dayan2007, raymer2022,Dorfman2016}. This corresponds to $\phi=\sigma_e /\sigma^{(2)}$. Using  $n=\phi A_0 /B$, under the condition that $A_e=A_0$, this criterion is equivalent to having the number of photons per spectral-temporal mode equal $f$. This conclusion agrees with expectation when recognizing that $f$ is approximately equal to 1. The understanding is that when the number of photons per spectral-temporal mode exceeds 1, we leave the isolated-pair regime.  Above the isolated-pair regime, photons from uncorrelated pairs overlap in time and generate excitation from TPA at rates comparable to ETPA, eliminating the benefit of temporal-spectral entanglement and reverting to quadratic flux scaling. At this flux, a transform-limited pulse of the same duration and intensity would have nearly the same ETPA efficiency \cite{raymer2022}. 

Unfortunately, it is unfeasible to observe the incident flux at which crossover occurs directly, due to the extremely low efficiency of TPA. However, the crossover point in the absence of spatial-coherence effects can be understood as a feature of the SPDC process and the acceptance bandwidth of the nonlinear process rather than being a feature unique to the TPA process. In this regard, the response of SFG is analogous to that of TPA. Because of this similarity, we use SFG to constrain experimentally the flux at which the crossover flux occurs for our SPDC (squeezing) source. To do this we measure directly the crossover that occurs in SFG using broadband entangled photons of variable flux. The agreement with predictions ensures that our theoretical description of the interaction captures the relevant dynamics for TPA, giving us confidence in our reported results for TPA in molecules.

\subsection{TPA rate with BSV}

Here we make a few more clarifying remarks comparing the TPA rate with bright squeezed vacuum to that with classical light. As noted above, Eq. (\ref{ETPA g2}), the TPA rate of an instantaneous two-photon process is proportional to the degree of second-order coherence, $g^{(2)}(0)$, of the driving field \cite{Spasibko2017_checkhova_g2g3g4, Boitier2011, Janszky1987, Louden2000book, iskhakov2012superbunched, raymer2022}. For classical coherent-state light arising from laser excitation we have $g^{(2)} (0) = 1$ and for squeezed vacuum $g^{(2)} (0) = 3$. For thermal-like light $g^{(2)} (0) = 2$. Values of $g^{(2)} (0)$ greater than 1 can be thought of as an increase in TPA rate arising from the quadratic nature of the interaction and the increased fluctuations in the driving field leading to higher probability for two or more photons present in a single temporal-spectral mode. However, in the case of interest the two-photon transition has a finite bandwidth, and thus the bandwidth and spectral correlations of the incident field are important considerations. In the limiting case of exceedingly narrow TPA linewidth, $\Delta \nu_f$, in comparison to the bandwidth of the field $B$, the incoherent (spectrally uncorrelated) portion of the field has an exceedingly small overlap with the transition, and contributes negligibly to the overall TPA rate \cite{Dayan2007,raymer2022}. On the other hand, the TPA rate arising from the coherent (correlated) portion of the field at high-gain is comparable to the TPA rate arising from classical light. In the case of a moderate TPA linewidth, the incoherent contribution is no longer negligible and the final TPA rate will lie somewhere between these two bounds. A more complete discussion of these effects can be found in \cite{raymer2022}.

\section{Experimental Overview}

Below, we describe our experimental results, which 1) validate the theoretical prediction that the TPA rate of BSV and coherent excitation are the same, to within a factor of 3, in Rhodamine 6G; 2) observe the crossover between linear and quadratic flux-scaling regimes in SFG as a means to characterize the state of our squeezed light source and as verification of our theoretical models and experimental techniques; 3) confirm that TPA of pulsed BSV with average fluxes comparable to maximal achievable entangled photon rates remain in the quadratic-scaling regime down to state-of-the-art detection thresholds; and 4) replicate recent experiments that report signatures of ETPA \cite{tabakaev2022} at lower detection threshold reduced background rates. The replication experiment demonstrates no detectable ETPA signal despite the reduced background and detection threshold in comparison with  \cite{tabakaev2022}.

Based on these results, which are in agreement with our theoretical predictions, we conclude that the established theory for ETPA is not missing any crucial effects that could explain the excessively large ETPA efficiencies reported elsewhere. Importantly, the benefit due to time-frequency correlations is not generally sufficient to overcome the low fluxes required to stay in the isolated EPP regime, even in well-suited samples such as Rhodamine 6G.

 \section{Experiments}

ETPA in Rhodamine dyes has been studied by several groups \cite{parzuchowski2020, Tabakaev2020, tabakaev2022, Villabona-Monsalve2017}. These dyes are appealing candidates for probing TPA with no resonant intermediate states (that is, only virtual-state pathways are available), due to their convenient absorption spectrum relative to available light sources, high fluorescence quantum yield, and ease of use arising from high solubility and stability. This allows for sensitive detection at highly concentrated solutions that can be optimized for fluorescence detection.

\subsection{Replication Experiment}

In a previous publication we reported bounds on the ETPA efficiency of Rhodamine 6G (R6G) for entangled photons generated from continuous-wave (CW) excitation via SPDC centered around a wavelength of 1064 nm \cite{Landes2021PRR}. Subsequent reports observed signals attributed to ETPA in R6G along with the linear and quadratic dual-scaling behavior characteristic of nonlinear interactions with entangled photons, as discussed above \cite{tabakaev2022}. This dual-scaling is a key feature of ETPA not previously observed in any other ETPA study. The reported non-zero ETPA fluorescence  detection rates (on the order of 10 counts/s after background subtraction of approximately 215 counts/s) \cite{tabakaev2022} did not contradict the bounds that had been set by our previous work\cite{Landes2021PRR}. That report motivated us to improve the threshold for detecting ETPA relative to our previous report and to attempt to detect ETPA signals under nearly identical conditions as in \cite{tabakaev2022}. 

We made every attempt to replicate the exact experimental conditions of \cite{tabakaev2022}, for example the optical beam parameters, lens focal lengths, interlens distances, squeezed-light source, detection geometry, etc..., while making minor modifications to improve detection sensitivity, as described below and in the supplemental information \cite{SI}. As mentioned in the Introduction, we used both a CW laser as in \cite{tabakaev2022} as well as a picosecond pulsed laser to generate squeezed vacuum in both low- and high-gain regimes, the latter allowing us to explore the regime of BSV-induced ETPA for comparison with the CW results and testing of theory predictions \cite{raymer2022}. The pulsed version of the experiment enabled clear detection of TPA-induced fluorescence. In the case of BSV excitation, we verified the presence of the TPA process by confirming the fluorescence emission lineshape \cite{SI} and quadratic dependence of fluorescence intensity with excitation flux. In addition, we used the pulsed source to confirm the theory's ability to predict properties of the SFG signal arising from squeezed vacuum, including observation of the crossover region at intermediate flux.

\begin{figure}
    \begin{center}
        \includegraphics[width = 8.5 cm]{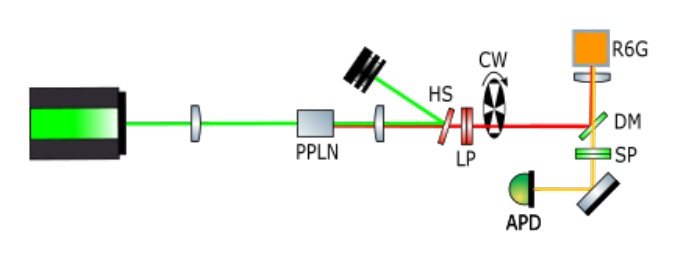}
    \caption{Experimental apparatus for the replication experiment. A prism separates any residual harmonics from the desired 532 nm CW laser light used to generate the entangled photon pairs at 1064 nm in a 20 mm periodically-poled lithium niobate (PPLN) crystal. After generation of the entangled pairs, the pump is separated using a harmonic separator mirror (HS). After this, the pairs travel through a chopper wheel (CW) with 50\% duty cycle. Finally, long-pass filters (LP) ensure that no pump light passes to the sample. A dichroic mirror (DM) reflects the pairs to the Rhodamine 6G sample (R6G). Fluorescence from the sample is collected in the backward geometry before passing through the dichroic mirror and a set of short-pass filters (SP), and being focused onto the detector (APD).}
    \label{fig: replication experiment}
    \end{center}
\end{figure}

Fig. \ref{fig: replication experiment} shows a schematic of the replication experiment. The pump generating the SPDC was a CW laser with a wavelength of 532 nm and bandwidth of approximately $\Delta \nu_{CW}=5$ MHz (\textit{Coherent Verdi V5}). A prism was used to separate any residual 1064 nm light generated by the laser. The beam was then focused into the down-conversion crystal with a 500 mm focal-length lens, resulting in an approximately 70 $\mu m$ beam waist. The down-conversion crystal was a 2 cm long magnesium-oxide doped periodically poled lithium-niobate (PPLN) crystal (\textit{Covesion MSHG1064-0.5-20}). The crystal was phase-matched for collinear, non-degenerate type-0 down-conversion via temperature tuning in a home-built oven. The SPDC light generated in this process was collimated using a 100 mm focal-length achromatic lens with anti-reflection (AR) coating optimized for 532 nm and 1064 nm (\textit{Edmund Optics \#33-211}). The pump was separated from the SPDC beam via a harmonic separator mirror (\textit{Newport  10QM20HB.21}). The SPDC beam then passed through a chopper wheel with 50\% duty-cycle operating at 400 Hz. Any remaining 532 nm light was then blocked by two long-pass filters (\textit{Thorlabs FELH0900, FELH0750}). 

After preparation of the entangled photons, the beam was routed to the `sample-and-detection' apparatus, which collected back-scattered light from the sample. The SPDC light was reflected off a dichroic mirror (\textit{Thorlabs DMSP0900}), and focused onto the sample using a 3 mm focal-length aspheric lens AR-coated for near-infrared (NIR) light (\textit{Thorlabs C330TMD-B}). The focal spot in the sample was near the diffraction limit so that the entanglement area was approximately equal to the spot size itself, that is  $A_e=A_0$, as assumed in our theoretical discussions above. The fluorescent sample was mounted on a linear translation stage enabling optimized placement in the longitudinal direction of the beam. The results reported in this paper utilized a standard 1-cm quartz cuvette with approximately 1-mm thick quartz window thickness. A thin sample cell using a microscope coverslip as a window, similar to that used in \cite{tabakaev2022}, was also tested with no significant differences in observed signal rates in both pulsed and CW experiments. Fluorescence emitted in the backward direction was collected and collimated by the focusing lens and transmitted through the dichroic mirror onto a filter stack (\textit{Thorlabs FESH0850, FESH0650, Semrock NF-03-532}). An 11 mm focal-length aspheric lens AR coated for visible wavelengths focused the fluorescence onto a free-space coupled avalanche photodiode (APD) operating in photon-counting mode (\textit{Laser Components Count-10B}). The dark rate of the detector was measured to be approximately 3 Hz, which is approximately 2 orders of magnitude smaller than that reported in \cite{tabakaev2022}. All elements of the experiment were light-shielded such that additional counts from stray light were not distinguishable from the intrinsic rate of dark counts of the detector alone. 

Alignment was optimized by maximizing the measured TPA fluorescence count rate from the sample when driven by BSV generated from the pulsed laser pumping the same crystal used to generate the CW low-flux SPDC, as detailed in the next section. The pulse-pumped BSV source was also used to validate the differential measurements used to suppress any slow drifts in background noise. The differential measurements is defined as the difference between detected counts in time intervals in which the beam was either allowed to pass through or completely blocked by a rotating shutter, or 'chopper wheel'. There were approximately 400 open/closed cycles per second. Validation was carried out both at high signal rates ($10^3$ to $10^5$ cps) as well as at fluorescence signal rates below 1 cps, yielding consistent results at all tested measurement durations, validated up to a maximum duration of 60,000 s. Background measurements under four different experimental conditions:--with the laser off; SPDC beam blocked; phase-matching temperature detuned; and on blank (without dye) ethanol solutions--all yielded no detectable signal above the dark counts. For a thorough discussion of the differential measurements see supporting information \cite{SI}.

\begin{figure}
    \centering
    \includegraphics[width = 8.5 cm]{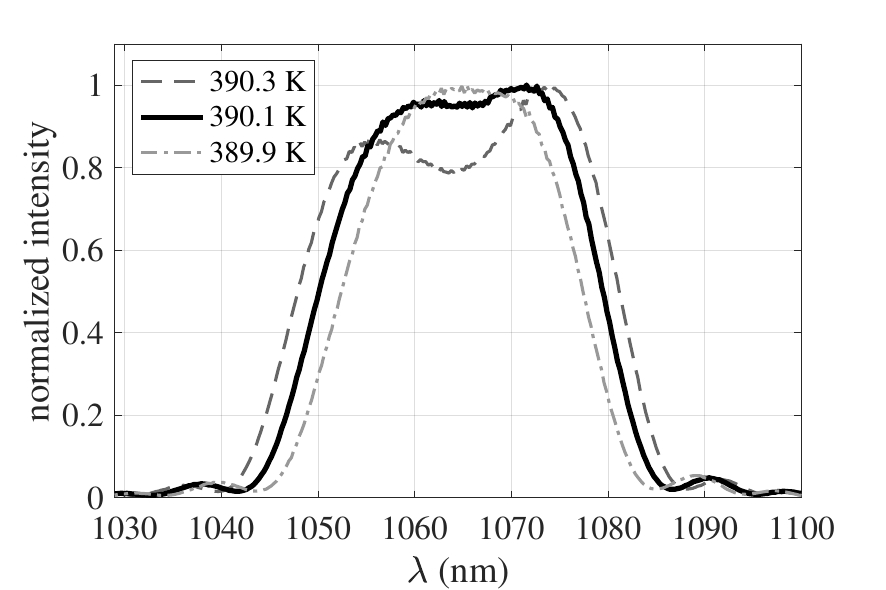}
    \caption{Marginal spectrum of the CW-pumped PDC source at various crystal temperatures. Spectra were measured on a 1000-m-long dispersive fiber CW time-of-flight spectrometer. Experiments presented in this paper were operated near a temperature of 390.1 K.}  
    \label{fig:cwpdcspec}
\end{figure}

The spectral characteristics of the CW SPDC at low gain were characterized via time-of-flight spectrometry. For this measurement the SPDC beam was coupled into a single-mode fiber via an 8-mm focal-length aspheric lens AR-coated for NIR (\textit{Newport KGA240-B-MT}). The fiber-coupled SPDC light was then split by a 50:50 fiber beam splitter (\textit{Thorlabs TW1064R5F2B}). One outgoing fiber was sent directly to a superconducting nanowire single-photon detector (SNSPD) (\textit{IDQuantique}), and the second outgoing fiber was sent through 1 km of optical fiber (\textit{Nufern 780HP}), which imparted the necessary optical dispersion for a time-of-flight measurement (converting frequency to time) before arrival at a second SNSPD. Due to the stochastic nature of the CW SPDC process, the photon pairs from the source arrive at the detector without reference to any external clock. However, the relative arrival time of each delayed photon contains spectral information due to the tight temporal correlation of the pairs \cite{avenhaus2009, BaekCW_TOF}. The measurement was calibrated using spectral filters with known sharp wavelength cutoffs. This measurement confirmed degenerate phase-matching around a central frequency of 1064 nm and a FWHM spectral bandwidth of 30 nm, as shown in Fig. \ref{fig:cwpdcspec}. 

To minimize the effects of drift and noise, we used `chopped' measurements of ETPA rates. The `chopped' measurement rates are defined as the difference between the detector counts measured when the SPDC beam is unblocked (open channel) by the chopper wheel and when it is blocked (closed channel). For a signal rate $R_S$ arising from fluorescence, the number of counts in a measurement time $T$ is expected to be $N_S=R_S T$ with uncertainty $\Delta N_S = \sqrt{R_S T}$. To determine whether a measured signal can be attributed to a process other than noise, we set the detection threshold by requiring the measured signal should be greater than 5 times the expected noise due to dark counts, that is we must have $N_S\geq 5 \Delta N_{noise}=5\sqrt{N_{noise}}= 5 \sqrt{D T}$. Here $D$ is the dark count rate (assumed constant). This allows us to define the detection threshold rate $R_{det}=5 \cdot 2 \sqrt{D/T}$. The factor of 2 accounts for the 50\% duty cycle of the chopped measurement scheme. For our experiments, $D=3$ Hz and $T=60,000$ s, giving $R_{det} = 0.07$ Hz.  Any signal with fluorescence rate above this detection threshold should be observable within our experiment. For comparison, note that the equivalent detection threshold rate in the previous experiment \cite{tabakaev2022} would be approximately $R_{det} \approx 10\sqrt{200/2\times10^4}=1$ Hz. 
All cited rates and powers are adjusted to account for the chopper duty cycle, and correspond to the average values while the beam is unblocked.

\subsection{Replication Result}

The PPLN crystal was pumped by a maximum CW power of 1 W, which generated approximately 200 nW of SPDC power measured after additional bandwidth filtering (\textit{Semrock 1055/70}). In order to facillitate comparisons to BSV experiments, we cite the total SPDC power within this spectral window, without attempting to adjust for pair-production efficiency. Shorter-duration measurements at higher powers were also conducted without observing evidence of TPA, however a power of 1 W was chosen to minimize temperature-dependent and photorefractive effects in the SPDC crystal. Measurements of fluorescence generated via TPA of entangled photons in a cuvette containing a 5-mM solution of Rhodamine 6G (\textit{Sigma Aldrich 252174}) in ethanol were made with a duration of 60,000 seconds, with no statistically significant difference between open and closed (chopper) channels. At this measurement duration, our detection threshold was 0.07 Hz (4.2 counts/minute). The measurement was repeated using a thin sample cell similar to the one used in \cite{tabakaev2022}. It was also repeated using a sample of the R6G dye that was used in \cite{tabakaev2022}, graciously provided by the authors to rule out sample-based discrepancies. No significant differences between samples or sample cells were observed in the repeated measurements with the CW or the pulsed configuration.

\subsection{Moderate-Gain BSV-Induced TPA}

Using the same PPLN crystal pumped by a pulsed laser, we investigated the behavior of TPA driven by the source at moderate gain. We then reduced the flux in an attempt to approach the low-gain regime with a source known to be capable of driving observable fluorescence from TPA of squeezed vacuum. 

The pulsed SPDC source utilized pump pulses with 532-nm center wavelength, which were about 8 ps in duration (\textit{Lumera Laser Hyperrapid 50}). The 8 ps pulse duration is far longer than the entanglement (correlation) time of the exciting light ($T_e=1/B=  \text{110 fs}$), making the theoretical models discussed earlier valid for our experiments. The repetition rate was varied between 100 kHz and 5 MHz. Otherwise the experimental apparatus was unchanged from the CW case, with the beam alignment to the CW beam confirmed by coupling into single-mode fiber as well as overlap on a CCD camera \textit{(Hamamatsu ORCA-ER-1394)} at the location of the crystal, and near the sample itself.

The BSV-induced signal can be unambiguously attributed to fluorescence from TPA, as verified by spectral measurements and quadratic scaling\cite{SI}. This is a major benefit of our current experiment, which no previous experiments in the ETPA literature have been able to achieve, due to the limited rates of detectable fluorescence. 

To approach the low-gain regime, the average power of the BSV generated by the pulsed source was kept constant at 300 nW, which is approximately the maximum attainable power in the CW regime. At 100 kHz laser repetition-rate, this provided a clearly measurable TPA signal of 88 Hz. From here, the repetition rate of the laser was increased incrementally, with the pump intensity adjusted to ensure the average power of the generated BSV remained constant. Increasing the laser repetition rate at constant average BSV flux reduces the number of photons per pulse, and the measurements are reported in terms of photons per pulse.

\begin{figure}
    \centering
    \includegraphics[width = 8.5 cm]{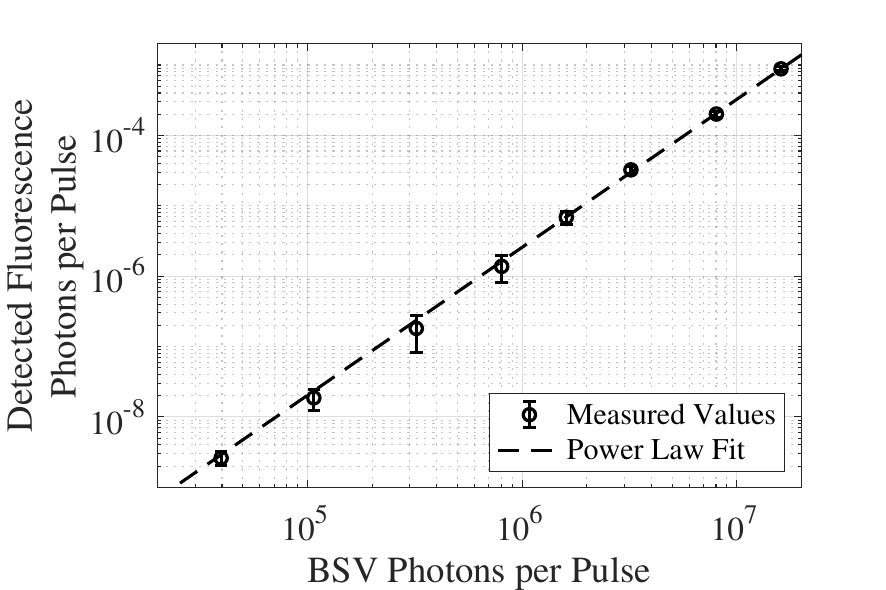}
    \caption{Double-logarithmic plot displaying measurements of TPA-induced fluorescence driven by pulsed BSV at an average power of 300 nW using a chopped measurement system. Error bars of 3$\sigma$ are shown. Quadratic scaling dependence (dashed line) is observed over the entire measurement (power law fit coefficient = 2.1). Repetition rates were varied from 100 kHz to 5 MHz at constant average power, and values are normalized to per-pulse rates---higher repetition rate corresponds to lower per-pulse flux.  Due to limitations of the experimental system, the lowest two data points were taken at 5 MHz at 100 nW and 37 nW BSV flux, respectively. The chopper wheel was omitted for these two data points, with open/closed channels being defined relative to the repetition rate of the laser. Measurement durations were adjusted to ensure $5\sigma$ threshold was met for all measurements.}  
    \label{fig: BSV}
\end{figure}

Due to the quadratic scaling of TPA intensity with peak BSV power, the rate of TPA is reduced as the repetition rate is increased, despite the average power remaining constant. This remains true until the linear scaling regime is reached, after which the repetition rate does not impact the detected fluorescence rate.  

The results of this experiment are shown in Fig. \ref{fig: BSV} in terms of per-pulse BSV intensities reported as average photon number per pulse. The quadratic dependence can be readily observed down to our minimum exposure flux. We were able to increase the repetition rate to 5 MHz while keeping the average BSV power constant at 300 nW. At higher repetition rates the pump power and thereby the BSV gain was insufficient to reach 300 nW of BSV. At 5 MHz repetition rate and 300 nW BSV power, TPA was still observable at a detection rate 0.9 Hz. Two further measurements were made at 5 MHz, with reduced average powers of 100 nW and 37 nW. These measurements took 4 hours, and 4 days respectively to reach the same statistical significance levels as previous measurements--indicative of the extremely unfavorable scaling behavior of  measuring weak, shot-noise-limited, quadratic signals. The last two measurements also omitted the use of the chopper wheel, and instead used the laser pulse-train to distinguish `open' and `closed' channels in order to maximize the signal-to-noise ratio of our measurement. 

The lowest-flux measurement was made with approximately 40,000 photon pairs per pulse, or approximately 400 pairs per spectral-temporal mode for our source. Notably, despite the benefit of being far above the isolated-pair regime---which is expected to be near 1 pair per spectral mode---at the same average powers as those attainable in the CW experiment, the fluorescence rate is well below that reported in \cite{tabakaev2022}. This result is in agreement with the null result from our CW experiment. 

A notable difference between our CW and pulsed experiments is the spectral bandwidth of the pump pulse, which determines the tightness of the spectral correlation between the entangled photon pairs \cite{Dayan2007, Dorfman2016}. This constitutes a significant experimental difference between pulsed and CW versions of the experiment, since the bandwidth of the CW pump is orders of magnitude narrower than that of the pulsed pump. This difference is not expected to affect the TPA efficiency in R6G, since the TPA linewidth is much broader than the pulsed-laser bandwidth, and thus the correlations present in the pulsed case are sufficient to maximize resonant overlap with the two-photon transition \cite{raymer2021tutorial}.

\subsection{Comparison of BSV and `Classical' Efficiencies}

To confirm another relevant point of our understanding of TPA arising from light generated via SPDC, we designed an experiment to compare the rate of TPA from a `classical'  (coherent-state) laser reference beam to the efficiency of TPA from BSV generated by our source. As discussed in the theory section, the relative efficiency for pulses of equal duration is expected to fall between 1 and 3, the former for narrow TPA linewidth and the latter for ultra-broad phase-matching\cite{raymer2022, Spasibko2017_checkhova_g2g3g4}.  

\begin{figure}
    \begin{center}
        \includegraphics[width = 8cm]{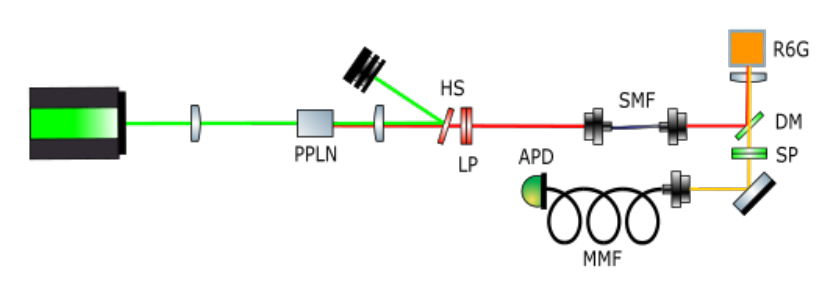}
    \caption{After generation of BSV in a 10 mm PPLN crystal, the BSV is separated from the pump via harmonic separator mirror (HS). Residual pump is blocked by a set of long-pass filters (LP). The BSV is then coupled into a connectorized 5-cm single-mode fiber stub (SMF). In the characterization step, the light coupled into the fiber stub was routed to SNSPDs for coincidence counting and time-of-flight spectrometry measurements. After characterization, and without adjusting the input coupling, the following experimental step was performed. The light was collimated out of the fiber stub and sent to a TPA detection apparatus in the backwards collection geometry. The pairs were reflected off a dichroic mirror (DM) to a 3-mm-focal-length aspheric lens which focused the pairs onto the R6G sample. Fluorescence was collected by the same lens, and passed through the same dichroic mirror (DM) before passing a set of short-pass optical spectral filters (SP). The fluorescence was then coupled into a multi-mode fiber (MMF), and sent to a single-photon avalanche diode (APD) for detection. This configuration was also used to confirm the spectral properties of the R6G fluorescence in high-gain BSV experiments. Separately, the classical reference beam was focused into the same fiber stub for the reference measurement. This ensures perfect spatial-mode matching between the coherent laser and BSV sources.}
    \label{fig: TPA BSV vs Classical_Exp}
    \end{center}
\end{figure}

To ensure identical spatial characteristics of the beams, a modified configuration was used that included a spatial-mode filter, as shown in Fig. \ref{fig: TPA BSV vs Classical_Exp}. This configuration used a 10-mm PPLN crystal (\textit{Covesion MSHG1064-1.0-10})  rather than the 20 mm PPLN crystal which was used in the above experiments. In this second configuration, the chopper wheel was omitted and the BSV beam was coupled into a 5-cm single-mode optical fiber stub (\textit{Nurfern 780HP}) using an 8-mm focal-length aspheric lens with NIR AR coating (\textit{Newport KGA240-B-MT }). The fiber stub was terminated in an FCPC connector. Its output could be coupled to a time-of-flight spectrometer, which verified the spectral and joint-spectral properties of the fiber-coupled BSV, as discussed in the supplemental information \cite{SI}.

\begin{figure}
    \centering
    \includegraphics[width = 8.5 cm]{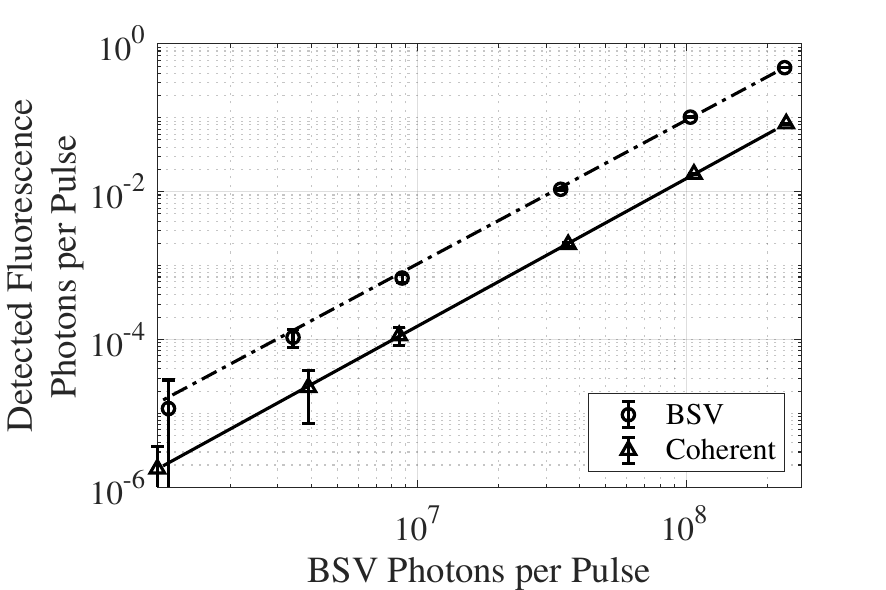}
    \caption{Comparison of scaling properties of fluorescence rate with BSV and coherent-state laser input power. Both show purely quadratic scaling, down to the measurement threshold limit. The relative efficiency of the BSV is 1.8 times as efficient as coherent excitation after accounting for pulse-duration effects. This is in good agreement with expectations. The duration of the 1064 nm laser pulse was measured to be 9.2 ps via intensity autocorrelation. The duration of the BSV pulse was measured to be 2.5 ps by the same measurement.}
    \label{fig: Scaling Comparison}
\end{figure}

Coupling into the fiber stub was optimized on coincidence counts in the low-gain SPDC regime. The spectral characteristics of the coupled light were also verified via measurement of the joint spectral intensity (JSI) of the photons coupled into the fiber. These were compared with previous JSI measurements\cite{SI}. After verifying the SPDC coupled into the short fiber displayed the desired JSI, the fiber going to the spectrometer was decoupled and light was collimated via an 8-mm-focal-length aspheric lens and routed to the detection apparatus in a backward collection geometry as described above. 

Fluorescence emitted in the backwards direction was collected and collimated by the focusing lens, and passed through the dichroic mirror through a filter stack (\textit{Thorlabs FESH0850, FESH0650, Semrock NF-03-532}), before being coupled into a multi-mode fiber, via an 8-mm aspheric lens (\textit{Newport KGA240-B-MT}). This multi-mode fiber was coupled to a spectrometer (\textit{Ocean Optics Flame}) to verify that the spectral properties corresponded to known properties of fluorescence from Rhodamine 6G.

The classical laser reference beam was coupled into the same fiber stub as the BSV, with all of the subsequent optics remaining unchanged, guaranteeing identical spatial properties (up to chromatic variations due to the broad BSV spectrum). 

To measure the temporal properties of the BSV, the beam after spatial mode filtering was routed to an autocorrelation experiment consisting of a Michelson interferometer prior to a 0.7-mm beta barium borate (BBO) crystal with broad phase-matching bandwidth. The pulse duration was used to calibrate the relative efficiency of the BSV compared to that of the classical reference laser. See further discussion in the supplemental information \cite{SI}.

The results of the efficiency comparison are summarized in Fig \ref{fig: Scaling Comparison}. Both sources showed quadratic scaling as confirmed by power-law fits of the data. After normalizing by the duration of the pulses, the relative TPA efficiency of the BSV was measured to be a factor 1.8 times as efficient as the classical reference laser. This result is in good agreement with expectations for this factor to be in the range 1 to 3 for a TPA sample with moderately broad TPA lineshape (resulting from optical dispersion in the system and the finite TPA molecular linewidth.) Crucially, no evidence of enhanced efficiency is observable when using BSV compared to classical laser light to drive TPA.

\subsection{Sum Frequency Generation}

The final experiment we describe uses sum frequency generation (SFG) as a model for TPA. The theories described in \cite{Dayan2007, raymer2021tutorial} predict TPA scaling behavior with the squeezed light flux that changes from linear to quadratic at the crossover flux, as described earlier. The crossover is predicted to occur at at photon rate of approximately 1 photon per temporal mode. This crossover rate is an important parameter for estimating the low-gain efficiency if the low-gain efficiency cannot be measured directly. The moderate-gain BSV results shown in Fig. \ref{fig: BSV} confirmed that this crossover for ETPA does not occur until below 400 photon pairs per mode from our source.

To constrain this expected behavior, which cannot be observed in our ETPA data for lack of signal, we use SFG as a model as it has similar dynamics to TPA\cite{Dayan2007}, with higher interaction strengths and thus higher and detectable efficiency.

This measurement was conducted using the 10-mm PPLN crystal  used for the BSV experiment, as well as the pulsed laser source operating at 10 MHz repetition rate. The chopper wheel was omitted, and no mode filtering was employed. The squeezed vacuum was collimated using a 200-mm focal-length achromatic lens. After spectral filtering (\textit{Thorlabs FELH0900, FELH0750}), the beam passed through a dispersion-compensating system (prism pair) to account for the second-order dispersion accrued in the optical system. The beam was then focused via a second achromatic, NIR AR-coated 200-mm focal-length lens into a nominally identical PPLN crystal under the same phase-matching conditions. The SFG acceptance bandwidth of the PPLN crystal was narrow compared to the bandwidth of the SPDC light driving the SFG, which ensures contributions primarily from the coherent contribution to the SFG signal.

\begin{figure}
    \centering
    \includegraphics[width = 8.5 cm]{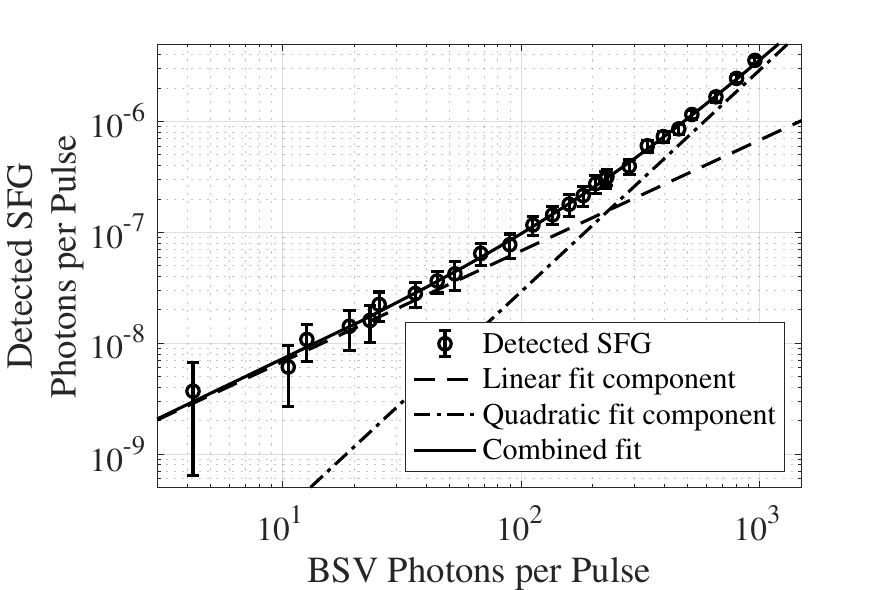}
    \caption{The crossover between linear and quadratic scaling regimes for SFG was measured to occur at approximately 250 photons per pulse, in good agreement with predictions of 1 photon per temporal mode. The measured result is higher by a factor of 2-3 than estimates of the number modes for the source, a difference attributed to spatial correlations present in the bulk SPDC process, which were not considered in the theoretical description.}
    \label{Scaling crossover in SFG. }
\end{figure}

After generation of the SFG beam, the remaining squeezed light was blocked by an interference filter stack (\textit{Thorlabs FESH0850, FESH0650, Semrock NF-03-532}), and the resulting light was coupled into a multi-mode optical fiber before detection on an APD (\textit{Laser-components COUNT-Blue 10-B}). The intensity of the BSV was measured via a multi-mode fiber-coupled power meter (\textit{Thorlabs S150C}). At low BSV intensities, the sensitivity of the power meter was limiting. The intensity of the pump was also monitored, to ensure that no errors were introduced due to measurement limitations of the power meter. 

Fig. \ref{Scaling crossover in SFG. } summarizes the result of this experiment. The crossover is measured to be near 250 photons per pulse, which is in good agreement with the predicted crossover for our source with ~100 temporal modes. No evidence of a large discrepancy in the predicted efficiency of the low-gain process for SFG was observed.

\section{Discussion}

Our experiment, which  had a substantially lower noise floor and thus lower detection threshold than the study reported in \cite{tabakaev2022}, did not reproduce the results reported there. At measurement durations up to 60,000 s we did not detect any signal distinguishable from dark-count noise using a CW source of entangled photons with 200 nW power. For measurements of this duration, utilizing the chopped experimental setup, we have a theoretical 5$\sigma$ detection threshold of 0.07 Hz. This methodology was validated by measuring a 0.9 Hz signal over the same duration generated from the pulsed experiment. In contrast, the results published in \cite{tabakaev2022} indicated an apparent ETPA count rate of approximately 9 Hz with incident power of 160 nW and otherwise near-identical conditions. 

Our experiment was run in several configurations: utilizing samples contained in a quartz cuvette  as well as in a thin glass cell similar to the one described in \cite{tabakaev2022}. Both were run utilizing a R6G sample purchased from Sigma-Aldrich (\textit{Sigma Aldrich 252174}) as well as with a sample of the same bulk stock used in \cite{tabakaev2022}, which was generously provided by the authors. This rules out differences due to sample inconsistencies or sample cell setup.

One source of discrepancy between the experiments is the choice of detector. The experiment in \cite{tabakaev2022} utilized a silicon APD with 500 $\mu$m active area, and approximately 200 Hz dark rate. Our experiment utilized an APD with 100 $\mu$m active area and 3 Hz dark rate, allowing for a lower detection threshold. The 25x difference in detector area between the two experiments is worth considering. Fluorescence collection is a difficult imaging problem, and non-optimal imaging could lead to overfilling of the detector. To verify that our experiment was not suffering from poor fluorescence collection, we imaged the fluorescence generated via TPA of BSV onto a CCD array (\textit{Hamamatsu ORCA-ER-1394}). The spot was seen to be tightly focused to ~70 ($\mu$m) FWHM spot on the array. Integrating the accumulated counts for a 100 $\mu$m x 100 $\mu$m spot and 500$\mu$m x 500$\mu$m spot, respectively, confirmed that the majority of the fluorescence was contained in the central spot on the CCD. Additionally, transverse translation of the APD itself yielded results consistent with  the fluorescence underfilling the APD, as desired.  

Another crucial difference between the experiments was the choice of anti-reflection coating for the final focusing and fluorescence collection lenses. The experiment in \cite{tabakaev2022} utilized an anti-reflection coating optimized for visible wavelengths (\textit{Thorlabs A-coating}) coating, which the manufacturer specifies at approximately 15\% reflection at 1064 nm at both the front and back face of the lens, which results in a significant reduction in pair-rate arriving at the sample, most of which is reflected directly back into the collection path. Our experiment utilized the same lens model, but with an anti-reflection coating optimized for NIR wavelengths (\textit{Thorlabs B-coating}), 'B'-coating which is specified at < 1 \% at 1064 nm. The collection losses at the central frequency of Rhodamine 6G fluorescence are specified as < 1\% for A-coating and approximately 7\% for the B-coating. Finally, we do not observe any background counts from a cuvette of ethanol alone. In \cite{tabakaev2022}, a 10 Hz background signal was observed from a blank cuvette of ethanol alone and was subtracted from their experimental runs. In contrast, in our experiment neither CW nor pulsed BSV---at significantly higher flux than possible in CW experiments---showed measurable signals from blank ethanol cuvettes. 

While we were not able to observe fluorescence from ETPA in the low-gain regime, we were able to observe TPA of BSV at similar average powers using pulsed, moderate-gain squeezed vacuum. In agreement with our low-gain null result, our measurements of TPA driven by moderate-gain squeezed vacuum showed quadratic scaling at all intensities down to approximately 400 photons per spectral-temporal mode. At these intensities the signal rates were far below those reported in \cite{tabakaev2022}, despite having the benefit of quadratic intensity scaling. This measurement would report on large increases in the TPA efficiency in the low-gain regime if they were present, as these would shift the expected crossover between linear and quadratic regimes, however no such shift was observed.

One major contribution of this particular experiment, absent in previous studies of ETPA, is that the BSV-induced measurements in this experiment can be unambiguously attributed to TPA-induced fluorescence by observing the spectral characteristics of the detected light, in addition to standard scaling properties, all of which could readily be measured in the same apparatus at high BSV fluxes.

To constrain further the measured value of the TPA crossover flux, we used SFG to study the characteristics of the squeezed vacuum field. In this measurement we were able to directly observe the crossover between linear and quadratic scaling behavior, which occurred at roughly 2.5 photons per mode. This is in satisfactory agreement with predictions of the theory. The minor discrepancy between predicted and observed crossover flux is possibly explained by weak spatial correlations present in the entangled photon beam, which are not considered in our model. Spatial correlations cannot explain the orders-of-magnitude differences observed in other ETPA experiments, and are expected to contribute minimally in tightly focused TPA experiments such as ours.

\section{Conclusions}

In spite of having a lower detection threshold for TPA, our results did not reproduce the results of the only ETPA experiment to date that has reported both the linear and the quadratic scaling with flux in ETPA. Our efforts at direct measurements of TPA at moderate and high gain, generated in the same experimental apparatus, yielded TPA rates well below those reported in \cite{tabakaev2022} despite relatively favorable conditions. Our combined results provide strong evidence that the orders-of-magnitude increases in TPA efficiency by using entangled light reported in \cite{Goodson2006,Goodson2017,Goodson2020Varnavski, Villabona-Monsalve2017,Tabakaev2020} cannot be  explained by the community's current understanding of ETPA, and may be resulting from some unexpected linear-optical affect mistakenly being attributed to ETPA. ETPA experiments that do not confirm the quadratic scaling of the process are particularly vulnerable to this kind of error \cite{Acquino2022, Cushing2022, Mikhaylov2022}. 

As we have shown theoretically in \cite{raymer2022} and experimentally in the present study, simply increasing the flux of entangled photons does not lead to greatly enhanced EPTA signals. When the flux is increased, the TPA process passes the crossover point, above which photons from uncorrelated pairs overlap. At this point the benefit of individual spectrally correlated pairs is lost, as the TPA efficiency is dominated by the quadratic scaling, which is comparable to the efficiency of `classical' (coherent-state) laser excitation. 

This is not to say that the phenomenon of ETPA does not exist. Clear, but minimal evidence of the effect was observed by Georgiades et. al.  in an atomic system where the narrow TPA linewidth gives a large cross section \cite{Georgiades1995}. There the TPA rate was seen to be increased by around a factor of two at extremely low fluxes as a result of the number correlations present in low-gain squeezing, although the narrow bandwidth of the squeezed light in that experiment precluded any study of enhancement by spectral correlation within photon pairs. The main barrier to detecting ETPA in solvated molecular samples is simply the fact that the TPA cross sections are so small as a result of the large homogeneous TPA linewidths. 

\section{Acknowledgements}    
We thank Andrew H. Marcus and Markus Allgaier for collaboration in earlier stages of this research. And we thank Rob Thew and Dmitri Tabakaev for providing complete information about their experiment and a sample of the dye they used. This work was supported by the National Science Foundation RAISE-TAQS Program (Grant No. PHY-1839216).

%\bibliography{References}

\end{document}

\preprint{APS/123-QED}

\title{Supplemental Information\\Limitations in Fluorescence-Detected Entangled Two-Photon-Absorption Experiments: Exploring the Low- to High-Gain Squeezing Regimes}% Force line breaks with \\
\author{Tiemo Landes}
\author{Brian J. Smith}%
\author{Michael G. Raymer}%
\date{\today}%
\maketitle

%\tableofcontents
\section{Spectrum of fluorescence driven by BSV}

\begin{figure}[H]
    \centering
    \includegraphics[width = 8.5 cm]{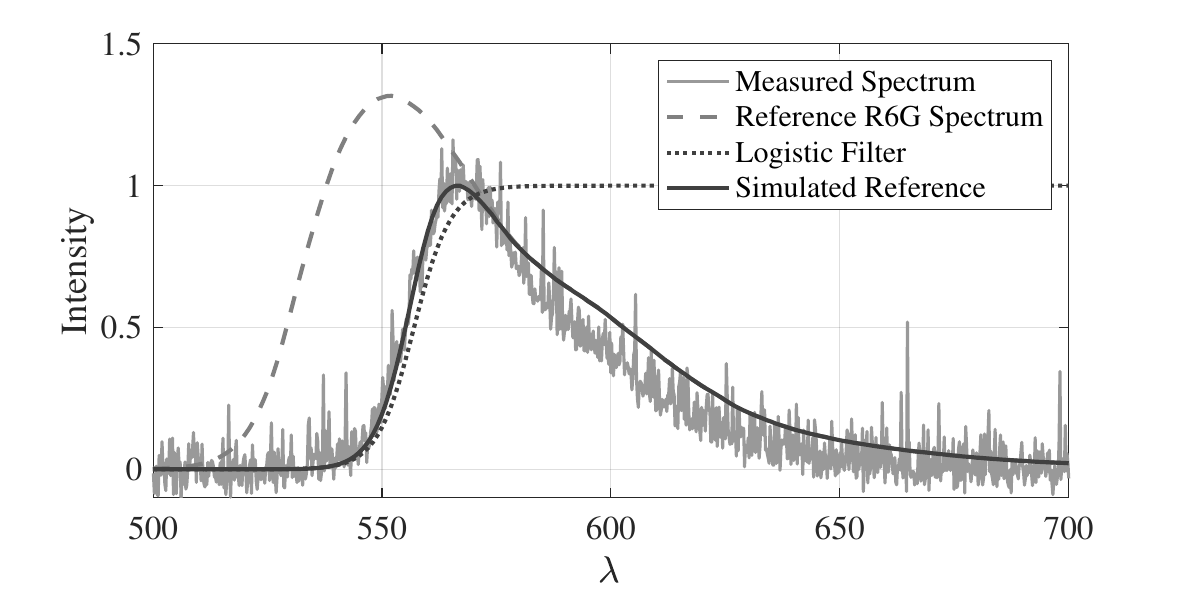}
    \caption{Comparison of measured R6G fluorescence spectrum driven by TPA of BSV spectrum with published reference fluorescence spectrum of R6G in ethanol \cite{dixon2005photochemcad} (dashed line).  After spectral filtering, fluorescence was measured with an average BSV power: 90 $\mu W$ at 100kHz repetition rate, and a measurement duration: 10 s (\textit{Ocean Optics Flame Spectrometer}). The dotted line shows an approximation of the spectral filters applied before the measurement was made. The solid line shows the product of the simulated spectral filter with the reference spectra. For ease of comparison all spectra are normalized to 1 at the peak of the measured spectrum, approximately 567 nm. Long pass filters approximated by logistic function centered at 557 nm with 4 nm steepness.}
    \label{fig: BSV fluorescence spectrum}
\end{figure}

To confirm that the measured signal corresponds to fluorescence driven by two-photon absorption of BSV, fluorescence from the collection apparatus was coupled into a multi-mode fiber, and measured on a spectrometer Fig. \ref{fig: BSV fluorescence spectrum}. After accounting for spectral filters, the fluorescence spectrum is in good agreement with the known fluorescence emission spectrum of Rhodamine 6G in ethanol. 

\section{Joint Spectral Measurements of Squeezed Vacuum}

\begin{figure}%[H]
    \centering
    \includegraphics[width = 8.5 cm]{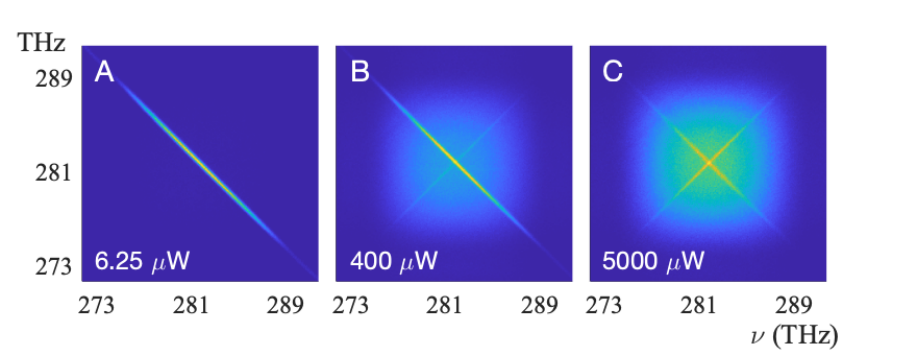}
    \caption{Joint Spectral Intensities of BSV source measured at varying pump power, showing the transition from tightly frequency anti-correlated photon pairs in low-gain SPDC, to high-gain BSV with many uncorrelated photons present. The pump powers used were: A) 6.25 $\mu W$,  B) 400 $\mu W$, C) 5000 $\mu W$.}  
    \label{fig: BSV_JSI}
\end{figure}

To characterize the SPDC generated in the pumped experimental setup, Joint Spectral Intensities (JSI) were measured at various levels of gain, Fig. \ref{fig: BSV_JSI}. The measurements were conducted using a time-of-flight (TOF) spectrometer (dispersive fiber and fast detectors), using two superconducting nanowire single-photon detectors (SNSPDs) to measure the frequency of individual pairs of photons generated via SPDC.  

The JSIs, Fig. \ref{fig: BSV_JSI}, show the transition behavior as the bright squeezed vacuum (BSV) transitions from low to high gain. In the low-gain limit, Fig. \ref{fig: BSV_JSI}A, the JSI for entangled pairs generated via SPDC can be seen. As the gain is increased, uncorrelated pairs are generated leading to a diffuse background, which can be observed in both the medium-gain Fig. \ref{fig: BSV_JSI}B, and high-gain Fig. \ref{fig: BSV_JSI}C measurements. At high gain, the intensity of the background term is large, and the diagonal feature has the same intensity as the anti-diagonal, as recently observed by another group using intensity correlations on conventional spectrometer \cite{Cutipa2022}.

JSIs were measured using SPDC generated from downconversion in a 10mm PPLN crystal. After spectral filters were applied, the light was coupled into a single-mode fiber. This was then passed through approximately 1000 m fiber spool before being split onto two separate SNSPDs. A pick-off from the pump beam was sent to a photodiode, and served as a timing reference for the JSI measurement. Medium and high-gain SPDC was attenuated using neutral density filters to prevent detector saturation. Average detection rates were kept at or below 1/5 of the laser's repetition rate to ensure unbiased spectral measurements. Spectral measurements were calibrated based with spectral band-pass filters with known frequency cutoffs.

\section{Spatial Mode Filter}

For measurements after coupling into the single-mode fiber stub, the fiber coupling of the BSV was optimized on the coincidence count rate coupled into the fiber stub. After optimizing the count rate, the JSI were measured to confirm that the coupled SPDC at low gain was centered at 1064 nm and was anti-correlated in frequency.

To compare the TPA efficiency of the BSV and classical reference, the temporal durations of the pulses need to be considered. The same 1064 nm pulsed laser is used both as classical reference and to generate the 532 nm SPDC pump laser. However because there are multiple nonlinear optical processes, the duration of the final BSV pulse is not the same as the original 1064 nm laser. In the course of the experiment, the duration of the Gaussian pulse is reduced in both the SFG process generating the 532 nm SPDC pump laser, as well as in the SPDC process generating the BSV. The duration of the BSV pulse is also a function of the gain in the BSV process, however, we make the simplifying assumption that this remains roughly constant over the experimental conditions we consider.

After spatial-mode filtering in the fiber stub, both the SPDC and classical reference were sent to a Michelson interferometer, prior to a 0.7 mm BBO crystal. Sum-frequency generation (SFG) of the light was used to measure the temporal duration of the pulse, to account for pulse-shortening effects in the SFG process. The classical reference pulse was measured to be 9.2 ps in duration, whereas the BSV pulse was measured to be 2.5 ps.

\section{Chopped Measurement Validation}
In order to rule out slow time drifts, TPA measurements at low signal rates were made using a "chopper wheel", which blocks and unblocks the SPDC beam many times a second. The chopper wheel has a 50\% duty cycle, and the open-closed frequency was set to approximately 400 Hz for all chopped measurements. By comparing the counts accumulated when the beam is blocked to those accumulated when the beam passes through the chopper wheel slots, the average signal and background rates are measured. Changes to background levels from sources not modulated by the chopper wheel are ruled out.
\begin{figure}%[H]
    \centering
    \includegraphics[width = 8.5 cm]{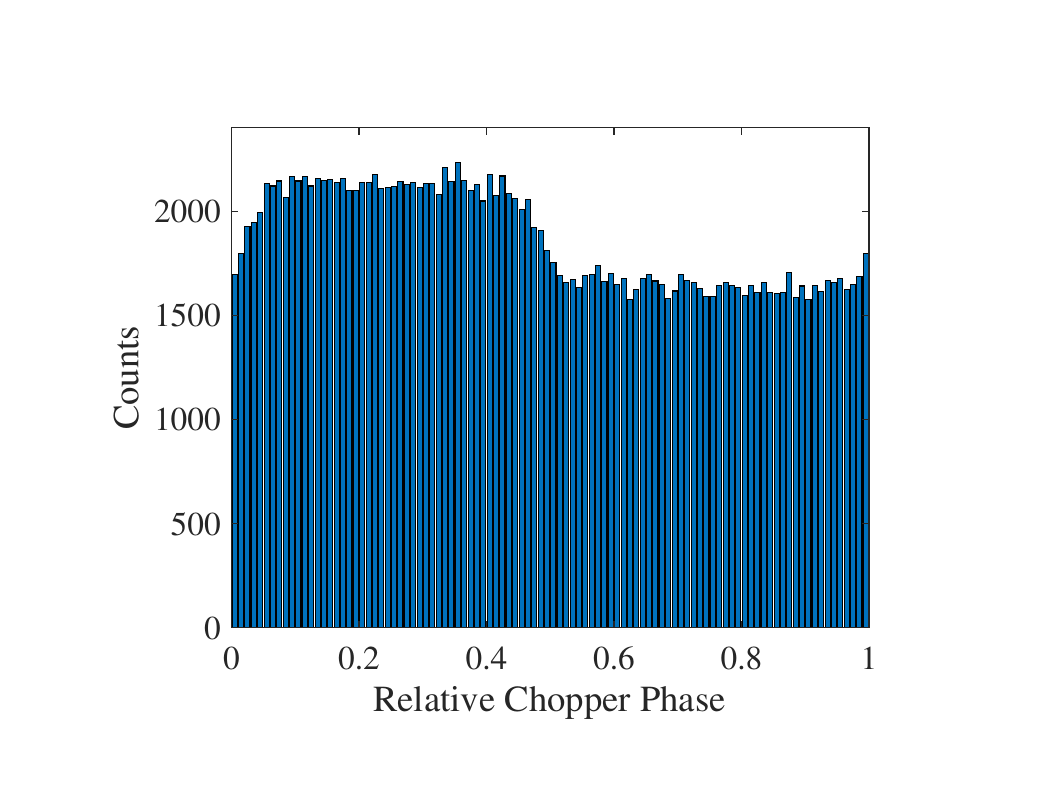}
    \caption{Histogram of event arrival times relative to the period of the chopper wheel's open-closed cycle. Data shown is from a measurement run with signal rate of approximately 1 Hz, generated via R6G fluorescence from TPA of BSV with a dark rate of approximately 3 Hz. The time intervals in which the chopper wheel is blocking and passing the beam can be clearly seen. Slight shoulders corresponding to partial beam blockage can be observed due to finite beam size.}
    \label{fig: Histogram}
\end{figure}
A "time tagger" (\textit{Qutools Qutau}) was used to correlate the arrival time of photon detection events to the position of the chopper wheel.  An example histogram taken at a detected fluorescence rate of approximately 1 Hz, with a 3 Hz background signal and is shown in Fig. \ref{fig: Histogram}.   

Detector dark events or dark noise has shot-noise-limited statistics, for which the variance of the number of detected events is equal to the number of detected events. This was confirmed by experimental observations. This allows for a straightforward calculation of the probability to measure a result as extreme as the one observed due to random fluctuations in the dark-noise signal. For a signal rate $R_S$ arising from fluorescence, the number of counts in a measurement time $T$ is expected to be $N_S=R_S T$ with uncertainty $\Delta N_S = \sqrt{R_S T}$. To determine whether a measured signal can be attributed to a process other than noise, we set the detection threshold by requiring a significant measured signal to be greater than 5 times the expected noise due to dark counts, that is we must have $N_S\geq 5 \Delta N_{noise}=5\sqrt{N_{noise}}= 5 \sqrt{D T}$. Here $D$ is the dark count rate (assumed constant). This allows us to define the detection threshold rate $R_{det}=5 \cdot 2 \sqrt{D/T}$. The factor of 2 accounts for the 50\% duty cycle of the chopped measurement scheme. 

\begin{figure}%[H]
    \includegraphics[width = 8.5 cm]{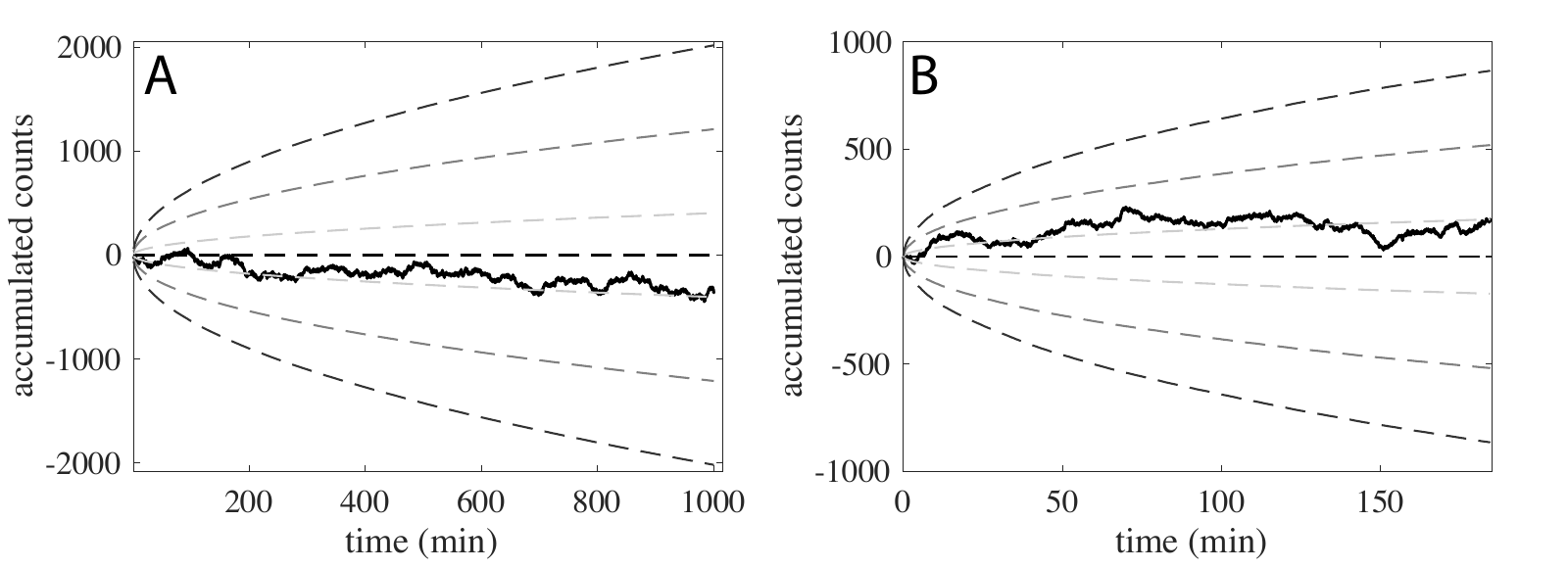}
    \caption{A) Measurement of fluorescence in 5 mM solution of Rhodamine 6G in ethanol, excited by TPA of SPDC generated by 1W of CW 532 nm pump laser. Average CW SPDC flux is 200 nW. The difference between open and closed channels after 60,000 second was -345 counts, well below the 2000 counts required to reach the 5$\sigma$ threshold. (The negative value indicates that more detection events occurred while the beam was blocked than while it was open). Dashed lines corresponding to significance levels $\pm \sigma$, $\pm 3 \sigma$, and $\pm 5\sigma$, are included as guides. The $5\sigma$ level at 60,000 s corresponds to 0.07 Hz. Note that the measured rate is twice the open-closed channel difference per time, due to the 50\% duty cycle of the measurement. B) Control measurement using a cuvette of ethanol with no dye added, measured using an average CW SPDC flux of 200 nW. The difference between open and closed channels after 10,000 seconds was 124 counts, well below the 822 counts required to reach the 5$\sigma$ threshold. The $5\sigma$ level at 10,000s corresponds to approximately 0.17 Hz.}
    \label{fig:Rhodamine 6G null}
\end{figure}

\begin{figure}%[H]
    \includegraphics[width = 8.5 cm]{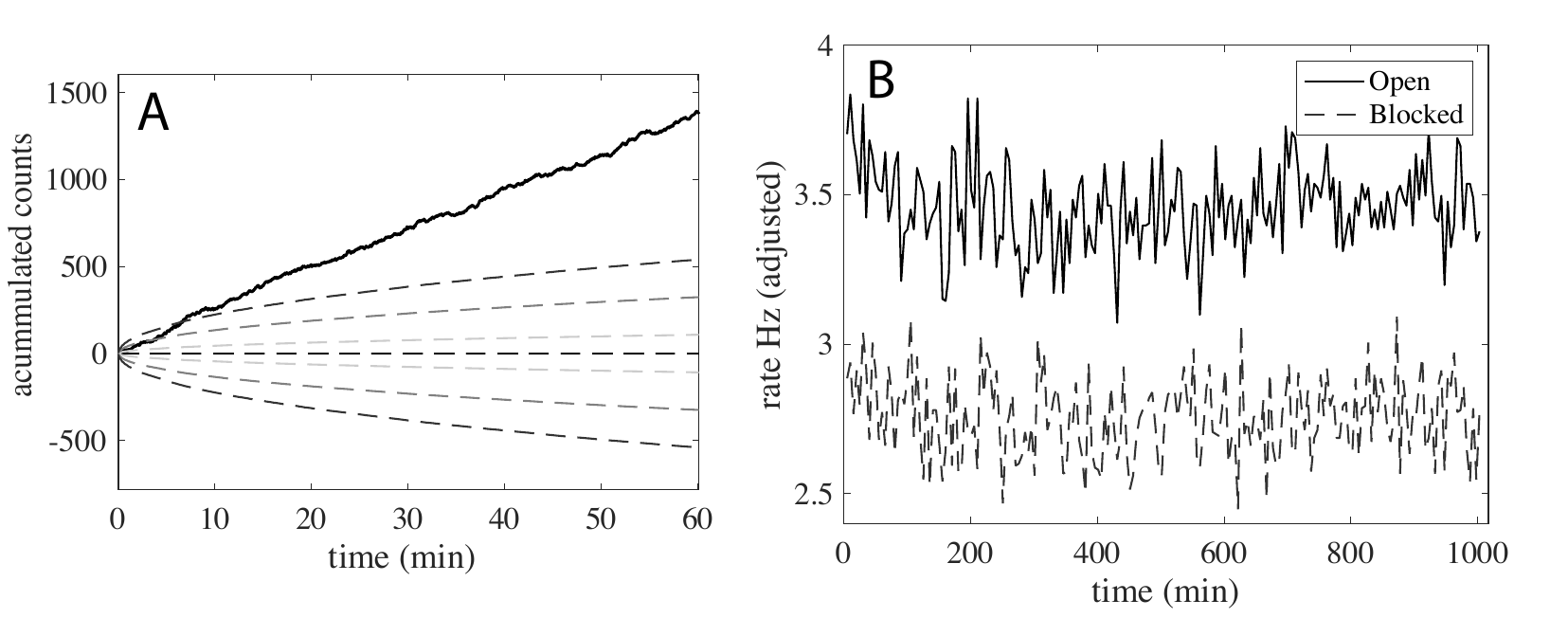}
    \caption{A) First hour of the long measurement is shown to highlight the difference between open and closed channels for chopped measurement of fluorescence excited by TPA BSV in Rhodamine 6G. Significance levels are shown for $\pm \sigma$, $\pm 3 \sigma$, and $\pm 5\sigma$. The $5\sigma$ confidence level surpassed after around 10 minutes. Note that the measured rate is twice the open-closed channel difference over time, due to the 50\% duty cycle of the measurement. B) Comparison of dark rate and signal rate over the course of a long measurement. Fluctuations in dark rate over time are mirrored in the signal rate, and absolute difference stays steady. Sub-Hz measurement stability is demonstrated over the entirety of a 60,000 s measurement.}
    \label{fig: R6G Validation}
\end{figure}

The difference between accumulated counts in the open and closed time-intervals for selected measurements are shown in Fig. \ref{fig:Rhodamine 6G null} and Fig. \ref{fig: R6G Validation}.

Fig. \ref{fig:Rhodamine 6G null} A shows a null result for a measurement of fluorescence from TPA of SPDC generated by CW excitation in Rhodamine 6G in ethanol over the course of a 60,000 second measurement. Fig. \ref{fig:Rhodamine 6G null}B shows a similar null result for TPA in a pure ethanol sample from the excitation conditions over the course of a 10,800 second measurement. In neither of these measurements does the signal level reach statistical significance, indicating that the measured signal cannot be differentiated from stochastic variations in the detector dark counts.

To confirm that the measurement was stable and capable of measuring signals at sub-Hz rates, the measurement was also carried out using Rhodamine 6G fluorescence generated TPA driven by BSV attenuated to the approximately 1-Hz level. This verifies the measurement's ability to measure weak signals over the course of many hours without drift or other issues. 

Fig \ref{fig: R6G Validation}A shows a similar trace over the first hour of the measurement. The count accumulation reaches the $5\sigma$ level within the first 10 minutes. Fig \ref{fig: R6G Validation}B shows the average rates in the open and closed time intervals over the entire duration of the measurement. This shows clearly that the difference between open and closed channels remains roughly constant over the course of the measurement, despite some drift in detector dark rate.

\newpage
%\bibliography{References}